# DESA: An R Package for Detecting Epidemics Using a School-Absenteeism Surveillance Framework

VINAY JOSHY
ZENY FENG
LORNA DEETH
KAYLA VANDERKRUK
JUSTIN SLATER

*Author affiliations can be found in the back matter of this article

## ABSTRACT

Absenteeism among elementary school children has proven useful for early detection of influenza epidemics within a population. This paper presents Detecting Epidemics from School Absenteeism (DESA), an R package developed to: (1) model epidemics using school absenteeism data, (2) raise alerts for potential outbreaks, (3) evaluate the timeliness of alerts using multiple metrics, and (4) simulate household populations, epidemics, and absenteeism patterns to support related research. DESA provides researchers and public health officials with a practical tool for improving early detection of seasonal influenza and other infectious diseases. The package is freely available on CRAN and GitHub for the R community.

**CORRESPONDING AUTHOR:**

**Zeny Feng**

Department of Mathematics and Statistics, University of Guelph, Canada

zfeng@uoguelph.ca

# (1) OVERVIEW

## INTRODUCTION

Surveillance systems play a vital role in identifying and controlling the spread of influenza, enabling public health authorities to implement timely interventions such as vaccination campaigns and antiviral distribution [1]. Traditional surveillance systems often rely on clinical reports and laboratory confirmed cases; however, these methods can be delayed and may not fully capture the real-time dynamics of an epidemic, particularly when a significant number of infected cases are not reported. To enhance the responsiveness of surveillance efforts, recent research has focused on incorporating alternative data sources, such as school absenteeism records, which can provide early indicators of community-wide epidemic arrivals [2, 3].

A significant advancement in this field was made by Vanderkruk et al. [4], who developed a comprehensive simulation study to evaluate school absenteeism-based surveillance methods. Their work, focused on the Wellington–Dufferin–Guelph Public Health Unit in Ontario, Canada, demonstrated the effectiveness of monitoring school absenteeism in detecting influenza epidemic onsets and introduced novel metrics for evaluating alert timing and accuracy. The simulation framework they created provided a foundation for exploring disease dynamics and testing surveillance strategies in a controlled environment.

This paper presents the `R` package DESA (Detecting Epidemics from School Absenteeism), which implements statistical models and evaluation metrics for early epidemic detection using school absenteeism data, extending the framework developed by Vanderkruk et al. [4]. The primary goal of DESA is to provide public health officials and researchers with a practical, accessible tool to detect the onset of epidemics through school absence patterns. While the original study was specific to one health unit, DESA generalizes the methodology to be applicable across different regions and populations.

DESA centers on implementing lag-logistic regression models that capture the temporal relationship between school absences and disease occurrence, allowing for early warning of incoming epidemics. The package includes multiple evaluation metrics to assess alert timeliness and accuracy, enabling users to select optimal model parameters based on their specific surveillance objectives. To support method development and validation, DESA also includes simulation capabilities for generating epidemic scenarios and corresponding absenteeism data. By bringing these tools to CRAN, we aim to make these analytical tools widely available to the research community, enabling further exploration and development of early warning systems for epidemic detection.

## IMPLEMENTATION AND ARCHITECTURE

DESA enhances computational efficiency in simulating epidemic and school absenteeism data for large populations by utilizing the stochastic SIR approach instead of individual-level models used in Ref. [4]. This method maintains the ability to capture realistic population dynamics and disease spread. It is particularly important for simulating epidemics in large populations, such as cities or regional municipalities. With no existing software available for such simulations, and given that school absenteeism data collection was interrupted during the COVID-19 pandemic due to multiple school closures of varying durations, DESA enables the generation of school absenteeism data to support research in this area.

### Epidemic detection model

To predict the onset of epidemics using school absenteeism data, a statistical model is needed that can capture both the temporal relationship between absences and disease occurrence as well as seasonal patterns. Ward et al. [5] suggested that distributed lag-logistic regression models, which included fixed effects for seasonality and a random effect for school year, are suitable to model the relationship between the probability of having at least one reported case and school absenteeism pattern. Following their suggestions, a regional-level seasonal mixed effects binary logistic regression model was employed, expressed as:

$$\text{logit}(\theta_{tj}) = \beta_0 + \sum_{k=0}^{l} \beta_{k+1} x_{(t-k)j} + \beta_{l+2} \sin\left(\frac{2\pi t^*}{T^*}\right) + \beta_{l+3} \cos\left(\frac{2\pi t^*}{T^*}\right) + \gamma_j \tag{1}$$

where $\theta_{tj}$ represents the probability of having at least one reported case on day $t$ for school year $j$. The predictor variables $x_{(t-k)j}$ are the mean daily absenteeism percentages with lag times ranging from 0 to $l$, where $l$ is the lag size that represents the maximum number of days of past absenteeism data to include in the model. To account for seasonal variations in influenza, trigonometric functions with sine and cosine terms are included, where $t^*$ denotes the calendar day of the year when $x_{tj}$ is observed and $T^* = 365.25$ days represents the length of a year. Additionally, $\gamma_j$ captures random effects specific to each school year $j$, which is assumed to follow a normal distribution $N(0, \tau^2)$. This random effect term serves a dual purpose: it addresses the intra-correlation among daily absenteeism and epidemic observations within a given school year while simultaneously allowing for heterogeneity across different school years [5].

In practice, this model is used to generate an alert on day $t$ of school year $j$ if the predicted probability of at least one laboratory confirmed influenza case, $\theta_{tj}$, exceeds a predetermined threshold $\theta^*$. Lag size $l$ and the threshold value $\theta^*$ are two tuning parameters whose optimal values are selected based on some criteria or performance evaluation metrics, which will be discussed in the following subsection. The model shown in Equation 1 is trained using historical data (laboratory confirmed cases and school absenteeism data from previous year and start of the current school year) and parameters are estimated. The school absences data is then used to predict $\theta_{tj}$ for each day. When $\theta_{tj} > \theta^*$, an alert is raised for the onset of an incoming epidemic [4]. Consequently, the first epidemic year serves solely as training data, and no alert is generated for this year as insufficient information is available for model fitting.

## Evaluation metrics of raised alerts

Unlike other evaluation metrics for statistical modeling such as mean square error, Akaike information criterion (AIC) or Bayesian information criterion (BIC), an epidemic alert evaluation metric must reflect the accuracy and timeliness of the alert raised by the model. Furthermore, such a metric will be used for selecting tuning parameters for the statistical model (Equation 1), as described in the previous subsection. The reference date for a seasonal influenza epidemic is defined as the date of the second laboratory confirmed case when the first two consecutive confirmed cases appear within a 7-day window [4]. This date indicates the rise in infected cases towards the peak of an epidemic, rather than the beginning of an influenza season. This reference date serves as a crucial benchmark for evaluating the timeliness of epidemic alerts. Any alert raised before the start of a school year and after the reference date is ignored. In addition, a high quality alert should be raised not too early for its relevance and not too late for its timeliness. In the following subsections, we will describe several evaluation metrics that we considered and implement in DESA.

### *Evaluation metrics*

Two evaluation metrics were introduced for assessing the performance of epidemic prediction models: the false alarm rate (FAR) and accumulated days delayed (ADD) [5]. These metrics were designed to measure the accuracy and timeliness of alerts generated by the epidemic detection models, respectively. The FAR and ADD metrics were defined involving definitions of true and false alerts. As suggested in Ward et al. [5], true alerts were those raised within 14 calendar days prior to the reference date and any alert raised prior to this period are false alerts. In addition, an optimal alert ($\tau_{opt}$) in Equation 3 was suggested to be the one raised 14 days prior to the reference date. With these notions, the FAR and ADD are defined as:

$$\mathrm{FAR}_j = \begin{cases} \frac{n_f}{n_f+1}, & \text{if a true alert was raised} \\ 1, & \text{if no true alerts were raised} \end{cases} \tag{2}$$

where $n_f$ was the number of false alerts raised in school year $j$ [5].

$$\mathrm{ADD}_j = \begin{cases} \tau_{opt} - \tau_{obs}, & \text{if a true alert was raised} \\ \tau_{max}, & \text{if no true alerts were raised} \end{cases} \tag{3}$$

where $\tau_{opt}$ represents the ideal number of calendar days for an alert to be raised before the epidemic reference date, and $\tau_{obs}$ denotes the number of calendar days prior to the epidemic reference day when the first actual alert for that season was issued. As suggested in Ward et al. [5], FAR is used as a primary tool for model comparison and ADD served as a tie breaker when two models have equal FAR. A model that generates alerts with smaller FAR and ADD is more accurate and more timely, respectively.

### *Alert Time Quality metrics*

The Alert Time Quality (ATQ) metric was developed by Vanderkruk et al. [4], to provide a more nuanced evaluation of alert timing. ATQ implements a continuous penalty function that increases as alerts deviate from the optimal timing. The metric for alert $i$ raised in year $j$ is defined as:

$$\mathrm{ATQ}_{ij} = \begin{cases} \left(\frac{\tau_{opt}-\tau_{ij}}{k\tau_{opt}}\right)^{2a}, & \text{if } \tau_{ij} \le \tau_{opt} \\ \left(\frac{\tau_{opt}-\tau_{ij}}{k\tau_{opt}}\right)^{a}, & \text{if } \tau_{opt} < \tau_{ij} \le (k+1)\tau_{opt} \\ 1, & \text{if } \tau_{ij} > (k+1)\tau_{opt} \end{cases} \tag{4}$$

where $\tau_{ij}$ is the number of calendar days before the reference date that the alert was raised, $k$ controls the width of the acceptable alert window, and $a > 0$ determines the rate at which the penalty increases. The asymmetric power terms ($2a$ vs. $a$) reflect that early alerts are penalized more heavily than late alerts. Following Vanderkruk et al. [4], the DESA package implements default values of $k = 1.5$ and $a = 2$. Users may adjust these parameters based on local context; increasing $k$ widens the acceptable window for regions and decreases the ATQ value, but results in less predictable epidemic patterns. Adjusting $a$ modifies the relative penalty for early versus late detection. A model that generates alerts with lower ATQ would be more accurate and timely [4].

### *Aggregate performance measures*

To evaluate overall model performance, finding optimal values of tuning parameters through a validation process using multiple years historical data, two summary statistics of ATQ values are used. The Average ATQ (AATQ) considers all alerts raised for a given year, say, year $j$:

$$\mathrm{AATQ}_j = \begin{cases} \frac{\sum_{i=1}^{n_j} \mathrm{ATQ}_{ij}}{n_j}, & \text{if an alert is raised in school year } j \\ 1, & \text{if no alerts are raised in school year } j \end{cases} \tag{5}$$

where $ATQ_{ij}$ represents the ATQ value for the $i$th alert raised in school year $j$, and $n_j$ is the total number of alerts raised in that year. This metric provides insight into the overall quality of all alerts generated by the model for a given year.

Since public health officials typically respond when the first alert is raised, the First ATQ (FATQ) metric was developed to specifically evaluate the timing of the first alert:

$$FATQ_j = \begin{cases} ATQ_{1j}, & \text{if an alert is raised in school year } j \\ 1, & \text{if no alerts are raised in school year } j \end{cases} \tag{6}$$

where $ATQ_{1j}$ is the ATQ value of the first alert raised in year $j$. This metric focuses on the criticalness of the first warning, disregarding subsequent alerts [4].

*Weighted metrics*

The effectiveness of epidemic prediction models typically improves as more historical data becomes available for training [4]. To account for this, weighted versions of AATQ and FATQ metrics are implemented. Each year's prediction is assigned a weight based on the amount of training data available:

$$w_j = \frac{\text{Number of years in epidemic prediction model for year } j}{\sum_{i=1}^{n} \text{Number of years in epidemic prediction model for year } i} \tag{7}$$

These weights are used to compute the Weighted Average Alert Time Quality (WAATQ) and Weighted First Alert Time Quality (WFATQ):

$$WAATQ = \sum_{j=1}^{J} w_j AATQ_j \tag{8}$$

$$WFATQ = \sum_{j=1}^{J} w_j FATQ_j \tag{9}$$

where $J$ is the total number of years being evaluated. These weighted metrics give more importance to predictions made with more training data, providing a more nuanced evaluation of model performance as the surveillance system matures.

Model selection is performed by choosing the model that minimizes the chosen evaluation metric. The choice between using AATQ, FATQ, or their weighted variants depends on the specific goals of the surveillance system and the relative importance placed on overall alert quality versus first alert timing [4].

The various evaluation metrics presented offer different perspectives on model performance. FAR and ADD provide measures of accuracy and timeliness respectively [5], while the ATQ-based metrics offer more nuanced evaluations that consider both aspects simultaneously [4]. In practice, model selection typically involves examining all metrics to understand different aspects of performance. For instance, AATQ might be prioritized when overall alert quality throughout the epidemic period is crucial, while FATQ could be more important when early intervention is the primary concern. The weighted versions, WAATQ and WFATQ, are particularly valuable for ongoing surveillance systems as they account for the natural improvement in model performance as more training data becomes available. Users of the DESA package can select different metrics based on their specific surveillance goals, with the package providing multiple options to optimize models according to any of these criteria.

## Simulation framework for model evaluation

DESA also provides simulation capabilities to facilitate model testing, validation, and performance assessment. These simulation tools allow users to generate realistic epidemic scenarios, population structures, and resulting absenteeism patterns under controlled conditions. By simulating epidemics with known characteristics, researchers can evaluate the performance of different parameter configurations without waiting for influenza-like outbreaks in the real world. The following subsections detail the simulation components implemented in the package, beginning with population structure generation and progressing through epidemic dynamics and absenteeism patterns.

*Epidemic simulation models*

A widely used epidemic model for infectious disease dynamics is the class of compartment models. One example is the Susceptible-Infected-Removed (SIR) framework first introduced by Kermack and McKendrick [6]. In this classical framework, a population of size $N$ is partitioned into three compartments: susceptible ($S$), infected ($I$), and removed ($R$). The framework describes the progression of individuals through these states, where $S(t)$ represents the number of susceptible individuals, $I(t)$ the number of infectious individuals, and $R(t)$ the number of removed individuals (either recovered with immunity or deceased), each at time $t$. The dynamics of this system are governed by the rates at which individuals transition between these compartments.

The transmission process, which governs the $S$ to $I$ transition, is fundamental to epidemic modeling. In deterministic SIR models, this transition is typically modeled using the mass action principle, where the rate of new infections is proportional to the product of susceptible and infectious individuals. However, this approach can be overly simplistic for real-world applications, particularly when studying disease spread at the community level [7].

Individual-level models (ILMs), as formalized by Deardon et al. [8], offer a more sophisticated approach to modeling disease transmission. ILMs incorporate spatial and demographic heterogeneity by modeling the probability of infection for each susceptible individual based on their characteristics and relationships with infectious individuals [9, 10]. The probability of susceptible individual $i$ becoming infected at time $t$ is given by:

$$P(S_i(t) \rightarrow I_i(t)) = 1 - \exp(-\Omega_i(t)) \quad (10)$$

where $\Omega_i(t)$ represents the overall infectious pressure on individual $i$ at time $t$, which can incorporate various spatial and demographic factors.

While ILMs provide a rich framework for modeling epidemics, they present significant computational challenges, particularly for large populations. The computational complexity grows quadratically with population size, as each susceptible-infectious pair must be evaluated at each time step. This limitation was evident in the original implementation by Vanderkruk et al. [4], where simulation times became prohibitive for realistic population sizes.

To address these computational constraints while maintaining model fidelity, DESA implements a stochastic SIR (sSIR) framework. This approach combines the mathematical tractability of compartmental models with the randomness inherent in real epidemic processes. The foundational SIR model was introduced by Kermack and McKendrick [6] and has been extended in numerous ways, including stochastic variations to incorporate randomness in transmission dynamics [7, 11]. The probability of new infections in the sSIR model is given by:

$$P(S(t) \rightarrow I(t)) = 1 - \exp\left(-\alpha \cdot \frac{I(t)}{N} - \varepsilon(t)\right) \quad (11)$$

where $\alpha$ represents the transmission rate, $I(t)$ is the current number of infectious individuals, $N$ is the total population size and $\varepsilon(t)$ is a small spark term, specified by the user, that accounts for unexplained infectious sources outside the modeled population. The number of new infections at each time step is then drawn from a binomial distribution with parameters $n = S(t)$ and $p = P(S \rightarrow I|t)$. This formulation provides a computationally efficient alternative to ILMs while preserving the stochastic nature of disease transmission [11, 12].

This hybrid approach allows DESA to efficiently simulate epidemics in large populations while capturing the essential features of disease spread, including the inherent randomness in transmission events and the temporal dynamics of the epidemic. The sSIR framework provides a balance between computational efficiency and model realism, making it particularly suitable for public health applications where rapid simulation and analysis are essential [13, 14].

*Population simulation*

The simulation of realistic population structures is fundamental to modeling epidemic spread within communities. The DESA package implements a hierarchical approach to population generation, based on Ref. [4], beginning with broad geographical divisions and proceeding to individual household compositions. The first step involves the simulation of catchment areas, which represent distinct geographical regions within the study area.

Without loss of generality, a study region is represented by an $a \times a$ unit square, subdivided into equal-sized catchment areas. The number of schools within each catchment area is generated by a suitable distribution function available in base `R` with user specified parameter(s), for example, the Poisson distribution with a specified mean parameter. This spatial structure provides the foundation for implementing the stochastic SIR model at the population level, while allowing for local variations in contact patterns and transmission dynamics within each catchment area. Rather than simulating separate SIR models for each area, the entire population interacts as a single system in the epidemic simulation, with the spatial organization primarily serving to structure realistic contact networks and population distributions.

Elementary school populations are then simulated for each catchment area. Similar to catchment areas, school sizes can be generated by a suitable probability distribution function with user specified parameter(s), allowing researchers to adopt the variability in school enrollment sizes specific to their region of interest.

The household simulation process is divided into two distinct phases: households with children and households without children. For households with children, the simulation accounts for:

- Parent type (single or coupled)
- Number of children per household
- Age distribution of children, particularly elementary school age

The distributions for each of these characteristics can be customized through user-specified probability functions, with default uniform distributions provided. Households with children are generated iteratively until the simulated elementary school populations are satisfied, ensuring consistency between household structures and school enrollments. Each elementary school-aged child is assigned to a school within their catchment area, creating the necessary links between household and school populations. For households without children, the simulation considers:

- Distribution of household sizes
- Regional proportion of households with and without children

The number of households without children is calculated proportionally based on the number of households with children. These households are also assigned to catchment areas. Finally, the simulation generates individual-level data, expanding the household-level information to create records for each person in the population. This includes:

- Individual identifiers
- Household affiliations
- School assignments (for elementary school children)
- Spatial coordinates within catchment areas

Spatial locations for all households are assigned using complete spatial randomness within their respective catchment areas. This spatial information is retained for potential use in disease transmission models that incorporate geographical proximity. While this population structure captures the spatial and demographic characteristics at the individual level, the epidemic simulation itself operates using the stochastic SIR framework introduced in Section 'Epidemic Simulation Models'. Rather than tracking each potential transmission event between individual pairs (as in an ILM), the sSIR model simulates the epidemic dynamics at the population level with probabilistic transitions between compartments. The detailed population structure established through this simulation process ensures realistic household and school compositions, which influence absenteeism patterns when the epidemic spreads. Individual attributes such as school assignment and household affiliation are used to determine absence behaviors once infection status is determined by the sSIR model, creating the link between population-level epidemic dynamics and observable surveillance data at schools.

### *Epidemic simulation*

After generating the population structure, the `simepi` function from the DESA package simulates epidemic spread using a stochastic SIR framework. This two-stage approach first establishes the demographic and spatial characteristics of the population, and then applies the epidemic model to this population structure to simulate disease transmission. The initial population serves as the substrate on which the epidemic unfolds, with individuals transitioning between susceptible, infectious, and removed states according to the stochastic rules of the SIR model. The simulation generates day-by-day records of new infections, transitions between disease states, and tracking of individual infection status throughout the epidemic period. The simulation process incorporates both the progression of the disease through the population and the practical aspects of disease surveillance, such as reporting delays and case confirmation, which are implemented through separate probability mechanisms as detailed below.

The stochastic SIR implementation operates in discrete time steps, with the probability of new infections at time $t$ given by Equation 11. The number of new infections at each time step is drawn from a binomial distribution:

$$I_{new}(t) \sim \text{Binomial}(S(t), P(\Delta S \to I|t)) \quad (12)$$

The transmission rate $\alpha$ in Equation 11 is specified by the user, which can be calibrated to achieve realistic epidemic patterns, typically resulting in 3%–11% of the population becoming infected over the course of the epidemic [4, 13]. The infectious period is set to 4 days by default, reflecting typical influenza progression [15].

To simulate real-world surveillance conditions, the package implements a reporting mechanism where only a fraction of cases are laboratory confirmed and reported to health authorities. The reporting process includes:

$$C_{report}(t) \sim \text{Binomial}(I_{new}(t), p_{report}) \quad (13)$$

where $p_{report}$ represents the reporting probability. Reporting delays are modeled using an exponential distribution:

$$\text{delay} \sim \text{Exp}(\lambda) \quad (14)$$

where $\lambda$ is the inverse of the average reporting delay. This creates a more realistic surveillance scenario where case confirmations arrive with variable delays after infection. For each simulated epidemic, a reference date is established to mark the start of the epidemic.

### *Absenteeism data simulation*

`compile_epi` function in the DESA package aims to capture absences resulting from the infection of influenza and absences resulting from other causes referred to as baseline absenteeism. Baseline absenteeism is estimated using historical data from periods starting from September and typically preceding seasonal influenza activity. In Ref. [4], a baseline absenteeism rate of 5% is used, meaning that on any given day, a student has a 5% probability of being absent due to reasons unrelated to influenza infection:

$$P(\text{absent}|\text{other causes}) = 0.05 \quad (15)$$

For students who become infected during the simulated epidemic, the probability of absence increases significantly. The package implements a 95% probability that an infected student will be absent on each day of their infectious period:

$$P(\text{absent}|\text{infected}) = 0.95 \quad (16)$$

The daily absence probability for student $i$ at time $t$ is therefore:

$$P(\text{absent}_{i,t}) = \begin{cases} 0.95 & \text{if student } i \text{ is infected at time } t \\ 0.05 & \text{otherwise} \end{cases} \quad (17)$$

Daily absenteeism data is generated through the following process:

1. For each day $t$, identify the currently infected elementary school students
2. Generate absences for infected students using $P(\text{absent}|\text{infected})$
3. Generate absences from other causes for students using $P(\text{absent}|\text{other causes})$
4. Aggregate absences at both the school and regional levels

The aggregation of absence data occurs at multiple levels:

- Individual schools: Daily absence counts and percentages
- Regional level: Mean daily absence percentage across all schools
- Separation of infection related and baseline absences for analysis

The final output for each day includes:

$$\%_{absent,t} = \frac{\sum_{s=1}^{S} \text{absent}_{s,t}}{\sum_{s=1}^{S} \text{enrolled}_s} \times 100 \quad (18)$$

where $S$ is the total number of schools in the region, $\text{absent}_{s,t}$ is the number of absent students in school $s$ on day $t$, and $\text{enrolled}_s$ is the total enrollment of school $s$.

This approach to simulating absenteeism data capture features observed in real school absenteeism-based surveillance systems while maintaining tractable computations for large populations.

## Overview of DESA

The package implements a methodology for evaluation of alerts raised given epidemic and school absenteeism data.

### *Model evaluation*

The `eval_metrics` function implements the framework detailed in the previous sections and assesses the performance of epidemic alarm systems across various lags and thresholds using school absenteeism data. It evaluates the following metrics:

- False Alarm Rate (`FAR`): Proportion of alarms raised outside the true alarm window.
- Added Days Delayed (`ADD`): Measures how many days after the optimal alarm day the first true alarm was raised.
- Average Alarm Time Quality (`AATQ`): Mean quality of all alarms raised, considering their timing relative to the optimal alarm day.
- First Alarm Time Quality (`FATQ`): Quality of the first alarm raised, based on its timing.
- Weighted versions (`WAATQ`, `WFATQ`): Apply year-specific weights to `AATQ` and `FATQ`.

`eval_metrics` also identifies the best model parameters (lag and threshold) for each metric. The output is a list with three main components:

- `metrics`: An object containing:
  - List of matrices of each metric (`FAR, ADD, AATQ, FATQ, WAATQ, WFATQ`) for all lag and threshold combinations.
  - List of best model and their predicted epidemic alert days for each metric
- `plot_data`: Plot object to visualize epidemic data and the best model for each metric
- `results`: Provides summary statistics

## Population and epidemic simulation

The simulation framework is based on the simulation method developed by Vanderkruk et al. [4] but improves on the efficiency of simulating large populations. It implements a hierarchical approach to population generation and epidemic simulation. The core functions of this framework are detailed below.

- `catchment_sim`: Simulates catchment areas using a default gamma distribution for the number of schools in each area. The `dist_func` argument allows for specifying other distributions.
- `elementary_pop`: Generates of elementary school enrollment and allocates students to catchment areas using a default gamma distribution. This function requires the output from `catchment_sim`. The `dist_func` parameter can be adjusted to accommodate alternative distributions.
- `subpop_children`: Simulates households with children, utilizing output from the function `elementary_pop`. `subpop_children` necessitates the specification of population parameters, including the proportion of coupled parents, the number of children per household type, and the fraction of elementary school-aged children. Users can define custom distributions for simulating parents, children, and ages.
- `subpop_noChildren`: Simulates households without children, based on the outputs from `subpop_children` and `elementary_pop`. This function requires the user to specify the proportions of various household sizes and the overall fraction of households that include children.

- `simulate_households`: Creates a list containing two simulated populations: households and individuals.

Epidemic simulations are conducted using the `ssir` function, which implements the stochastic SIR model on a given population. Users can specify various aspects of the simulation, including:

- Duration of the epidemic period
- Initial number of infected individuals
- Transmission rate ($\alpha$ in Equation 11)
- Proportion of cases reported
- Number of simulated epidemics

The `ssir` function outputs an object of the S3 class, providing compatibility with summary and plot methods for analysis and visualization of the simulated epidemic data. For a complete list of function arguments, refer to the package manual.

*Data compilation*
After the epidemic data is simulated, `compile_epi` function compiles and processes epidemic data, simulating school absenteeism using epidemic and individual data as described in Section 'Absenteeism Data Simulation'. It creates a data set for actual cases, absenteeism, laboratory confirmed cases, true alarm windows, reference dates for each epidemic year and seasonal lag values.

## USAGE

DESA provides functions that use school absenteeism data to raise an alert for the arrival of an epidemic in a given population. For demonstration purposes, we illustrate a complete workflow using data simulated by other functions of DESA, from generating a population through epidemic detection and evaluating an alert for the generated epidemic. The example shows how to simulate a regional population with realistic demographic structure, model multiple epidemic scenarios, and evaluate various metrics. For reproducibility, the seed for the pseudo-random number generator is set to 656.

### Population simulation

We first begin by installing DESA through CRAN and then simulating 16 catchments of 80 × 80 unit area with the number of elementary schools each catchment area contains by a normal distribution of mean 3 and standard deviation of 1.

```
R> install.packages(DESA)
R> library(DESA)

R> set.seed(656)
R> catchment <- catchment_sim(16, 80,
       dist_func = stats::rnorm, mean = 3, sd = 1)
```

The enrollment size of elementary schools are generated from a gamma distribution with parameters $\alpha = 7.86$ and $\beta = 0.032$.

```
R> elementary<- elementary_pop(catchment,
                               dist_func = stats::rgamma,
                               shape = 7.86, rate = 0.032)
```

Population of households with children are simulated according to user specified parameters via the function `subpop_children`. Usually, these parameters are provided by demographic information from population censuses. We set `prop_parent_couple` as 0.77, which is the proportion of housholds with children that have two parents. A couple has a probability distribution of 0.36, 0.43, and 0.21 for having 1, 2, and 3+ school-age children (`prop_house_Children`), respectively. A single parent has a probability distribution of 0.58, 0.31, and 0.11 (`prop_children_lone`), respectively. The proportion of the children of elementary school age (`prop_elem_age`) is set as 0.53. `subpop_children` requires the output from `elementary_pop`.

```
R> house_children <- subpop_children(
       elementary, prop_parent_couple = 0.77,
       prop_house_Children = c(0.36, 0.43, 0.21),
       prop_children_lone = c(0.58, 0.31, 0.11),
       prop_elem_age = 0.53)
```

Population of households without children are simulated using demographic information as well. We set the proportions of households with 1, 2, 3, 4, and 5+ members (`prop_house_size`) as 0.23, 0.34, 0.16, 0.16, 0.09 and the proportion of households with children (`prop_house_Children`) as 0.43 respectively.

```
R> house_noChild <- subpop_noChildren(
       house_children, elementary,
       prop_house_size = c(0.23, 0.35, 0.17, 0.16, 0.09),
       prop_house_Children = 0.43)
```

The resulting population of 199,669 individuals represents a realistic demographic structure with proper household compositions and school assignments, as described in Section 'Population Simulation'. `simulate_households` combines the household simulations and generates individual-level data. Individual data can be extracted from the output of `simulate_households`, this will also give us the total number of individuals simulated within the population.

```
R> households <- simulate_households(house_children,
                                     house_noChild)
R> individuals <- households$individual_sim

nrow(individuals)

R> [1] 199669
```

### Epidemic simulation

Given the simulated population, we can simulate epidemic replicates using the `ssir` function. These replicates can be used to mimic epidemic historical data collected for a population over multiple years, and they are used to fit epidemic detection models. Here, we have simulated 10 epidemics, each with a time period set to 300 days. The mean epidemic start time (`avg_start`) is set to be 45 days after the seasonal influenza starts at day 0, with a minimum start date of 20 days. We set the infectious period (`inf_period`) to be 4 days, with a total of 32 individuals initially infected (`inf_init`) at day 0. The reporting delay (`lag`) is set to 7 days, and 2% of infected cases are reported (`report`). The transmission rate of the disease (`alpha`), as in Equation 11, is set to 0.298.

```
epidemic <- ssir(
    N = nrow(individuals), T = 300, alpha = 0.298,
    avg_start = 45, min_start = 20, inf_period = 4,
    inf_init = 32, report = 0.02, lag = 7, rep = 10)
```

Using the generic function `summary` on the `epidemic` object shows us that on average, 79,347.4 individuals were infected with an average of 1585 reported cases over 10 replicates. The epidemic simulation uses the stochastic SIR model described in Section 'Epidemic Simulation', with parameters chosen to reflect typical influenza transmission patterns. The summary statistics show that approximately 30% of the population becomes infected over the course of the epidemic, with only 2% of cases being reported. Figure 1 depicts a plot of the epidemic which is accessed via 'plot(epidemic)'.

```
R> summary(epidemic)

SSIR Epidemic Summary (Multiple Simulations):
Number of simulations: 10

Average total infected: 79347.4
Average total reported cases: 1584.5
Average peak infected: 5808.4

Model parameters:
$N
[1] 259905

$T
[1] 300

$alpha
[1] 0.298

$inf_period
[1] 4

$inf_init
[1] 16

$report
[1] 0.02

$lag
[1] 7

$rep
[1] 10
```

### Alert system evaluation

School-absenteeism is simulated using `compile_epi` function, where epidemic and individuals' data are required. The output is a data frame containing simulated absenteeism data with relevant information. For example in the data set (`absent_data`) generated below, where on a given school year (`ScYr`), say year 1, date (`Date`) of an epidemic runs from 1 to 300. On each day, there is an average reported percentage of absent students (`pct_absent`) and the total count of absent students in the population (`absent`). The `absent_sick` column represents the simulated absent counts due to infection of influenza on each day, `new_inf` represents the number of new infections per each day, and `reported_cases` represents the number of cases reported per each day. The `Case` column is a binary indicator of whether there have been at least one newly infected case, `sinterm` and `costerm` values are seasonal trigonometric terms, `window` and `ref_date` are binary indicators for 'true alarm' periods and reference dates respectively. `lag0` to `lag15` represent the lagged absenteeism values for each day.

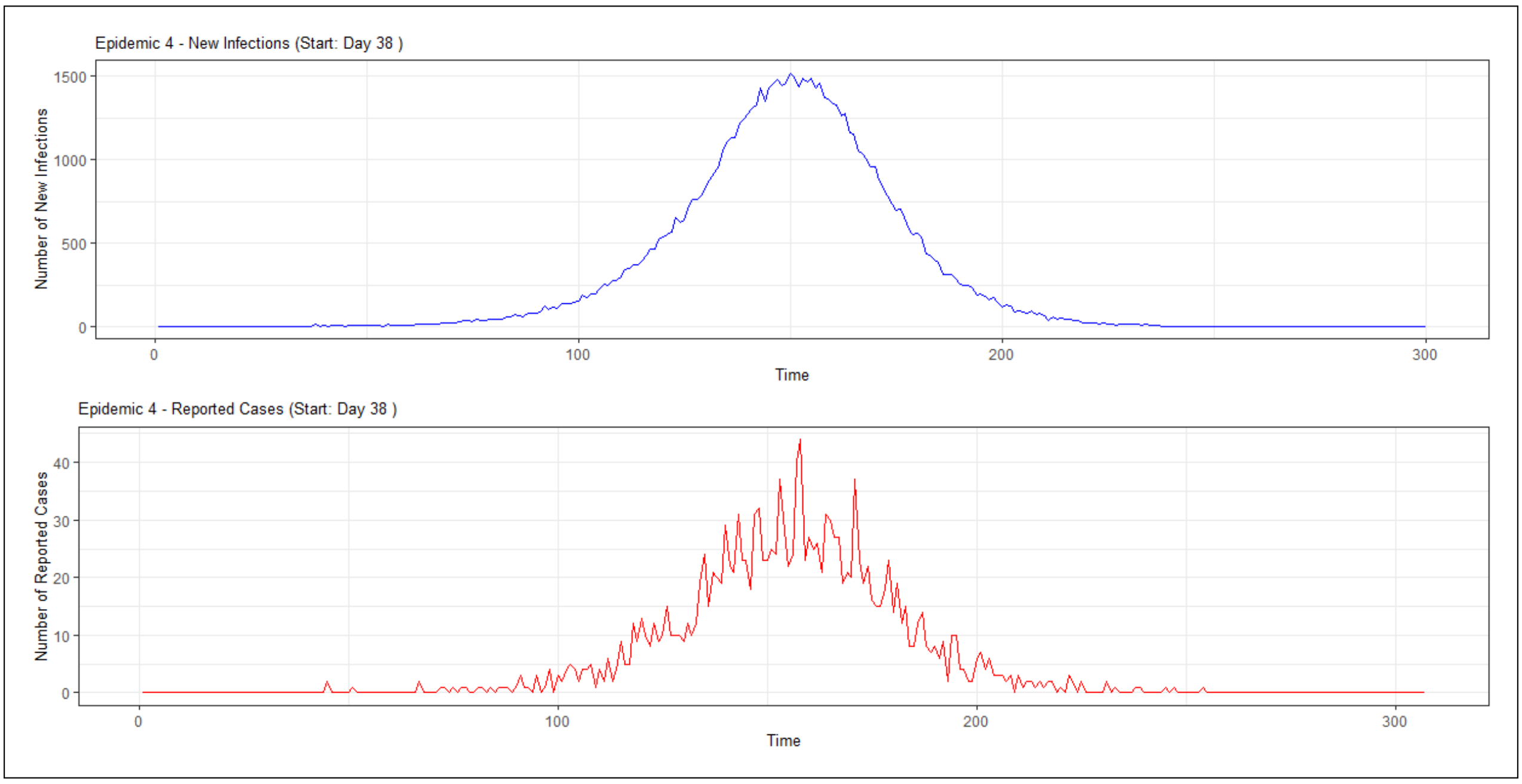

**Figure 1** Visualization of simulated epidemic dynamics from year 4, generated using `set.seed(656)` as specified in Section 'Population Simulation'. The top panel shows the number of new infections over time, demonstrating the characteristic epidemic curve with a peak of approximately 1500 daily cases around day 150. The bottom panel shows the corresponding reported cases after applying the 2% reporting rate and 7-day reporting delay, resulting in more variable daily counts with a maximum of 44 reported cases. The epidemic starts on day 38, showing an initial slow growth phase followed by exponential growth and eventual decline.

```
absent_data <- compile_epi(epidemic, individuals)
R> dplyr::glimpse(absent_data)

Rows: 3,000
Columns: 28
$ Date          <int> 1, 2, 3, 4, 5, 6, 7, 8, 9, 10, 11, 12, ...
$ ScYr          <int> 1, 1, 1, 1, 1, 1, 1, 1, 1, 1, 1, 1, ...
$ pct_absent    <dbl> 0.04969159, 0.04653166, 0.05107497, 0.0...
$ absent        <dbl> 1016, 972, 1058, 1058, 1037, 996, 1017, ...
$ absent_sick   <dbl> 0, 0, 0, 0, 0, 0, 0, 0, 0, 0, 0, 0, 0...
$ new_inf       <dbl> 0, 0, 0, 0, 0, 0, 0, 0, 0, 0, 0, 0, 0...
$ reported_cases <dbl> 0, 0, 0, 0, 0, 0, 0, 0, 0, 0, 0, 0, 0...
$ Case          <dbl> 0, 0, 0, 0, 0, 0, 0, 0, 0, 0, 0, 0, 0...
$ sinterm       <dbl> 0.01720158, 0.03439806, 0.05158437, 0.06...
$ costerm       <dbl> 0.9998520, 0.9994082, 0.9986686, 0.99763...
$ window        <dbl> 0, 0, 0, 0, 0, 1, 1, 1, 1, 1, 1, 1, 1, 1...
$ ref_date      <dbl> 0, 0, 0, 0, 0, 0, 0, 0, 0, 0, 0, 0, 0...
$ lag0          <dbl> 0.04969159, 0.04653166, 0.05107497, 0.05...
$ lag1          <dbl> NA, 0.04969159, 0.04653166, 0.05107497, ...
$ lag2          <dbl> NA, NA, 0.04969159, 0.04653166, 0.051074...
$ lag3          <dbl> NA, NA, NA, 0.04969159, 0.04653166, 0.05...
$ lag4          <dbl> NA, NA, NA, NA, 0.04969159, 0.04653166, ...
$ lag5          <dbl> NA, NA, NA, NA, NA, 0.04969159, 0.04653 ...
$ lag6          <dbl> NA, NA, NA, NA, NA, NA, 0.04969159, 0.04...
$ lag7          <dbl> NA, NA, NA, NA, NA, NA, NA, 0.04969159, ...
$ lag8          <dbl> NA, NA, NA, NA, NA, NA, NA, NA, 0.04969 ...
$ lag9          <dbl> NA, NA, NA, NA, NA, NA, NA, NA, NA, 0.0 ...
$ lag10         <dbl> NA, NA, NA, NA, NA, NA, NA, NA, NA, NA, ...
$ lag11         <dbl> NA, NA, NA, NA, NA, NA, NA, NA, NA, NA, ...
$ lag12         <dbl> NA, NA, NA, NA, NA, NA, NA, NA, NA, NA, ...
$ lag13         <dbl> NA, NA, NA, NA, NA, NA, NA, NA, NA, NA, ...
$ lag14         <dbl> NA, NA, NA, NA, NA, NA, NA, NA, NA, NA, ...
$ lag15         <dbl> NA, NA, NA, NA, NA, NA, NA, NA, NA, NA, ...
```

We can now evaluate the different metrics for each epidemic year using `eval_metrics` function. The lag values for the lag-logistic regression models range from 1 to 15 and threshold values are from 0.1 to 0.6 by 0.05 increments. Output of `eval_metrics` is a list of objects that contain raw and visualization data, on which functions like `summary` and `plot` can be used. Note that the function `eval_metrics` uses default ATQ parameters $k = 1.5$ and $a = 2$ from Equation 4, which can be modified via the `atq_k` and `atq_a` arguments.

```
R> alarms <- eval_metrics(absent_data, maxlag = 15,
                          thres = seq(0.1,0.6,by = 0.05),
                          atq_k = 1.5, atq_a = 2)
R> summary(alarms$results)

Alert Metrics Summary
=====================

FAR :
  Optimal lag: 3
  Optimal threshold: 0.45
  Minimum value: 0.2046
ADD :
  Optimal lag: 1
  Optimal threshold: 0.1
  Minimum value: 0

AATQ :
  Optimal lag: 5
  Optimal threshold: 0.25
  Minimum value: 0.0911

FATQ :
  Optimal lag: 3
  Optimal threshold: 0.4
  Minimum value: 0.1698

WAATQ :
  Optimal lag: 5
  Optimal threshold: 0.25
  Minimum value: 0.0886

WFATQ :
  Optimal lag: 3
  Optimal threshold: 0.4
  Minimum value: 0.1371

Reference Dates and Model Selected Alert Dates:
=====================

   year ref_date FAR ADD AATQ FATQ WAATQ WFATQ
1     1       20  NA  NA   NA   NA    NA    NA
2     2       51  37  37   37   37    37    37
3     3       45  31  31   31   31    31    31
4     4       38  38  24   34   38    34    38
5     5       43  41  29   34   41    34    41
6     6       61  48  47   47   48    47    48
7     7       50  43  36   43   43    43    43
8     8       45  36  31   32   36    32    36
9     9       81  67  67   67   67    67    67
10   10       45  45  31   32   33    32    33
```

The results show that the optimal performing lag values range from 1 to 5 days across different metrics, and optimal threshold values vary by metric, from 0.1 for ADD to 0.45 for FAR. The `Alert Metrics Summary` show the optimal lag and threshold values that produce the smallest value of each metric. Here, these optimal lag and threshold values are in combinations that minimize each respective alert metric. The `summary` function transforms the raw data `(alarms$results)` to provide the optimal lag and threshold value for each metric, reference dates for each epidemic year, and the model-selected alert dates for each metric. Model-selected alert dates represent the first day when an alert would be raised using the parameters that optimize each specific metric. Note that model-selected alert dates are NA for the first year because no data is available for training the epidemic detection model on this year, and the first year data is used solely for training the epidemic detection model for the later years.

```
R> alarmPlots <- plot(alarms$plot_data)
```

Plots for each epidemic year can be generated using the plot class on the plot object, `alarms$plot_data`. This will generate a list of plots corresponding to the different epidemic years. They can be accessed through `R` list indexing, one such plot is shown in Figure 2.

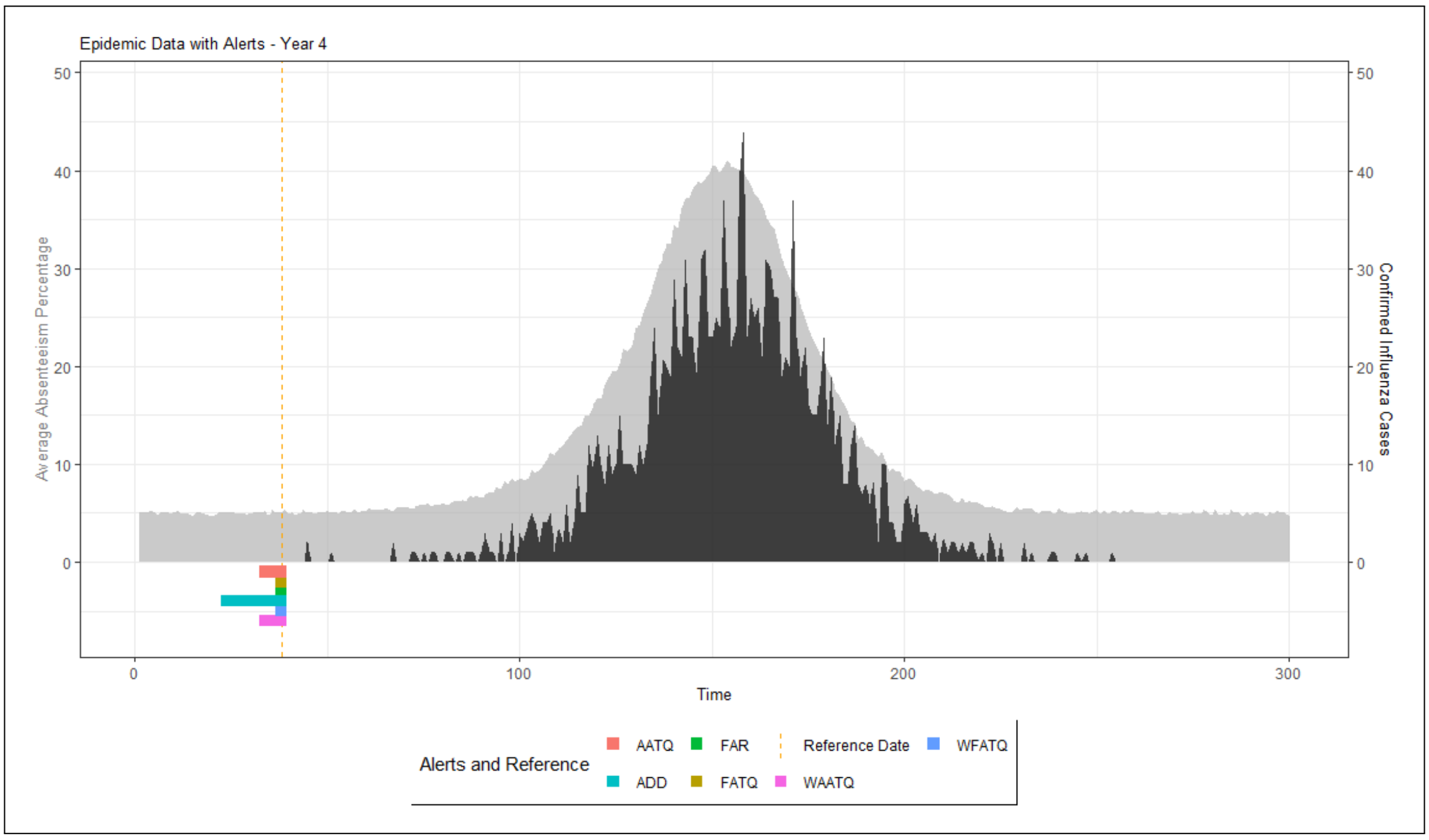

**Figure 2** Comparison of epidemic indicators and alert timings for year 4, generated using `set.seed(656)` as specified in Section 'Population Simulation'. The grey area shows daily school absenteeism percentage, while the black area represents laboratory confirmed cases (2% of true cases). The vertical dashed line indicates the reference date (epidemic start), and coloured squares below show the timing of alerts generated by different metrics. The clustering of alerts from different metrics shortly before the reference date suggests good agreement between methods, while their placement before the major increase in cases demonstrates the system's early warning capability. Notably, both weighted metrics (WAATQ, WFATQ) generate alerts at similar times to their unweighted counterparts, indicating robust detection regardless of weighting scheme.

### QUALITY CONTROL

The latest stable release of DESA is available on CRAN at https://CRAN.R-project.org/package=DESA. All functions were tested on simulated data and validated using real-world datasets during package development. Comprehensive unit tests, implemented using the `testthat` framework, ensure the correctness and reliability of the package's core functions. The package successfully passed all CRAN checks (`R CMD check`) without warnings, errors, or notes. A vignette included with the package demonstrates a complete workflow, allowing users to verify functionality through reproducible examples. A development version is also available on GitHub at https://github.com/vjoshy/DESA, supporting transparency, issue tracking, and community contributions.

## (2) AVAILABILITY

### OPERATING SYSTEM

The package was tested on Mac OS X, Windows and Linux.

### PROGRAMMING LANGUAGE

R version 4.1.0 or higher is required to install the package.

### ADDITIONAL SYSTEM REQUIREMENTS

No addiitonal requirements neccessary.

### DEPENDENCIES

R package dependencies: dplyr, purrr, zoo, ggplot2, testthat (≥3.0.0), gridExtra, rlang, scales.

### LIST OF CONTRIBUTORS

Vinay Joshy, Zeny Feng, Lorna Deeth, Justin Slater, Kayla Vanderkruk

### SOFTWARE LOCATION

**Archive**

***Name:*** CRAN
***Persistent identifier:*** https://doi.org/10.32614/CRAN.package.DESA
***Licence:*** GPL (≥ 3.0)
***Publisher:*** Vinay Joshy
***Version published:*** 1.0.0
***Date published:*** 07/05/2025

**Code repository**

***Name:*** Github
***Persistent identifier:*** https://github.com/vjoshy/DESA
***Licence:*** GPL-3.0 License
***Date published:*** 07/05/2025

### LANGUAGE

English

## (3) REUSE POTENTIAL

The DESA R package implements an epidemic detection model and quality evaluation tools for alerts raised by the detection model. It extends the simulation framework developed by Vanderkruk et al. [4] for a school absenteeism epidemic surveillance system. DESA provides computationally efficient and accessible tools, enabling researchers and public health officials to explore various aspects of early epidemic detection through stabilized surveillance principles [16] and simulation.

DESA provides researchers with the flexibility to investigate different surveillance scenarios through its modular design. Users can modify population structures, specify parameters for simulating populations, epidemics, and absenteeism data, and choose evaluation metrics to meet their specific research needs. This adaptability, combined with the package's computational efficiency, makes it particularly valuable for studying how school absenteeism-based surveillance systems might perform under various conditions before implementation in real-world settings. The current framework's applicability to other infectious diseases beyond influenza presents another avenue for future research [17].

Through its availability on CRAN, the DESA package aims to facilitate further research into school absenteeism-based surveillance systems and contribute to the development of more effective early warning systems for epidemic detection. Any issues, suggestions or improvements can be reported via Github or contacting corresponding authors.

## COMPETING INTERESTS

The authors have no competing interests to declare.

## AUTHOR CONTRIBUTIONS

Vinay Joshy: Software, Methodology, Writing – Original Draft; Zeny Feng: Methodology, Supervision, Writing – Review & Editing; Lorna Deeth: Methodology, Supervision, Writing – Review & Editing; Kayla Vanderkruk: Methodology, Data Curation, Writing – Review & Editing; Justin Slater: Methodology, Supervision, Writing – Review & Editing.

## AUTHOR AFFILIATIONS

**Vinay Joshy** orcid.org/0009-0003-7083-1256
Department of Mathematics and Statistics, University of Guelph, Canada

**Zeny Feng** orcid.org/0000-0002-8113-3619
Department of Mathematics and Statistics, University of Guelph, Canada

**Lorna Deeth**
Department of Mathematics and Statistics, University of Guelph, Canada

**Kayla Vanderkruk**
Unity Health Toronto, Canada

**Justin Slater** orcid.org/0000-0002-6648-5990
Department of Mathematics and Statistics, University of Guelph, Canada

## REFERENCES

1. **Thompson WW, Shay DK, Weintraub E, Brammar L, Cox NC, Anderson LJ, Fukuda K.** Mortality associated with influenza and respiratory syncytial virus in the United States. *JAMA*. 2004; 289(2):179–186. DOI: https://doi.org/10.1001/jama.289.2.179
2. **Viboud C, Boëlle PY, Carrat F, Valleron AJ, Flahault A.** Prediction of the spread of influenza epidemics by the method of analogues. *American Journal of Epidemiology*. 2004; 158(10):996–1006. DOI: https://doi.org/10.1093/aje/kwg239
3. **Lenaway DD, Ambler A.** Evaluation of a school-based influenza surveillance system. *Public Health Reports*. 1995; 110(3):333–337. Available at: https://www.ncbi.nlm.nih.gov/pmc/articles/PMC1382129/
4. **Vanderkruk KR, Deeth LE, Feng Z, Trotz-Williams LA.** ATQ: Alert time quality, an evaluation metric for assessing timely epidemic detection models within a school absenteeism-based surveillance system. *BMC Public Health*. 2023; 23(850):1–12. DOI: https://doi.org/10.1186/s12889-023-15747-z
5. **Ward MA, Stanley A, Deeth LE, Deardon R, Feng Z, Trotz-Williams LA.** Methods for detecting seasonal influenza epidemics using a school absenteeism surveillance system. *BMC Public Health*. 2019; 19(1232): 1–9. DOI: https://doi.org/10.1186/s12889-019-7521-7
6. **Kermack WO, McKendrick AG.** A contribution to the mathematical theory of epidemics. *Proceedings of the Royal Society A*. 1927; 115(772):700–721. DOI: https://doi.org/10.1098/rspa.1927.0118
7. **Keeling MJ, Rohani P.** *Modeling infectious diseases in humans and animals*. Princeton: Princeton University Press. 2008. DOI: https://doi.org/10.2307/j.ctvcm4gk0
8. **Deardon R, Brooks SP, Grenfell BT, Keeling MJ, Tildesley MJ, Savill NJ, Shaw DJ, Woolhouse MEJ.** Inference for individual-level models of infectious diseases in large populations. *Statistica Sinica*. 2010; 20(1):239–261.
9. **Rahul CR, Deardon R.** Individual-level models of disease transmission incorporating piecewise spatial risk functions. *Spatial and Spatio-temporal Epidemiology*. 2024; 50:100664. DOI: https://doi.org/10.1016/j.sste.2024.100664
10. **Deeth LE, Deardon R.** Latent conditional individual-level models for infectious disease modeling. *International Journal of Biostatistics*. 2013; 9(1):1–19. DOI: https://doi.org/10.1515/ijb-2013-0026

11. **Britton T.** Stochastic epidemic models: A survey. *Mathematical Biosciences*. 2010; 225(1):24–35. DOI: https://doi.org/10.1016/j.mbs.2010.01.006
12. **Allen LJS.** An introduction to stochastic epidemic models. In: *Mathematical Epidemiology*. Berlin: Springer; 2008. p. 81–130. DOI: https://doi.org/10.1007/978-3-540-78911-6_3
13. **Tokars JI, Olsen SJ, Reed C.** Seasonal incidence of symptomatic influenza in the United States. *Clinical Infectious Diseases*. 2018; 66(10):1511–1518. DOI: https://doi.org/10.1093/cid/cix1060
14. **Ferguson NM, Cummings DAT, Fraser C, Cajka JC, Cooley PC, Burke DS.** Strategies for mitigating an influenza pandemic. *Nature*. 2006; 442(7101):448–452. DOI: https://doi.org/10.1038/nature04795
15. **Lau LLH, Cowling BJ, Fang VJ, Chan KH, Lau EHY, Lipsitch M, Cheng CKY, Houck PM, Uyeki TM, Peiris JSM, Leung GM.** Viral shedding and clinical illness in naturally acquired influenza virus infections. *Journal of Infectious Diseases*. 2010; 201(10):1509–1516. DOI: https://doi.org/10.1086/652241
16. **Henning KJ.** What is syndromic surveillance? *MMWR Supplements*. 2004; 53:5–11.
17. **Unkel S, Farrington CP, Garthwaite PH, Robertson C, Andrews N.** Statistical methods for the prospective detection of infectious disease outbreaks: A review. *Journal of the Royal Statistical Society: Series A*. 2012; 175(1):49–82. DOI: https://doi.org/10.1111/j.1467-985X.2011.00714.x